\documentclass{PoS}

\newif\ifppn
\ppntrue

\usepackage{graphicx}
\usepackage{xcolor}

\newcommand{\BlackHat}{{\sc BlackHat}}

\newcommand{\SHERPA}{{\sc SHERPA}}

\def\Wj{$W\,\!+\,1$}

\def\Wjjjjj{$W\,\!+\,5$}
\def\Wjjjjjj{$W\,\!+\,6$}

\def\Wmjjjjj{$W^-\,\!+\,5$}

\def\Wjnp1{$W\,\!+\,(n\!+\!1)$}

\def\pT{p_{\rm T}}

\def\HTpartonic{{\hat H}_{\rm T}}

\title{High multiplicity $W$+jets predictions at NLO}

\ShortTitle{W+jets at NLO}

\author{\speaker{D.~Ma\^{\i}tre}\\
        Institute for Particle Physics Phenomenology, University of Durham, Durham DH1 3LE, UK\\
        E-mail: \email{daniel.maitre@durham.ac.uk}}

\author{Z.~Bern\\
        Department of Physics and Astronomy, UCLA, Los Angeles, CA 90095-1547, USA\\
        E-mail: \email{bern@physics.ucla.edu}}

\author{L.~J.~Dixon\\
        SLAC National Accelerator Laboratory, Stanford University, Stanford, CA 94309, USA \\
        E-mail: \email{lance@slac.stanford.edu}}

\author{F.~Febres Cordero\\
        Departamento de F\'{\i}sica, Universidad Sim\'on Bol\'{\i}var,  Caracas 1080A, Venezuela\\
        E-mail: \email{ffebres@usb.ve}}

\author{S. H\"oche\\
        SLAC National Accelerator Laboratory, Stanford University, Stanford, CA 94309, USA\\
        E-mail: \email{shoeche@slac.stanford.edu}}

\author{H.~Ita\\
        Physikalisches Institut, Albert-Ludwigs-Universit\"at Freiburg,
       D--79104 Freiburg, Germany\\
        E-mail: \email{harald.ita@physik.uni-freiburg.de}}
\author{D.~A.~Kosower\\
        Institut de Physique Th\'eorique, CEA--Saclay,
          F--91191 Gif-sur-Yvette cedex, France\\
        E-mail: \email{david.kosower@cea.fr}}
\author{K.~J.~Ozeren\\
        Department of Physics and Astronomy, UCLA, Los Angeles, CA 90095-1547, USA\\
        E-mail: \email{ozeren@physics.ucla.edu}}

\abstract{
\ifppn
\rule{1cm}{0cm}\vspace{-20cm}\\\hbox{\rm\small
B/F/421--13$\null\hskip 4cm \null$
IPPP/13/62$\null\hskip 4cm \null$
SLAC--PUB--15709
}
\hbox{\rm\small $\hskip 2.3cm \null$ UCLA--13--TEP--105
$\null\hskip 3cm \null$
IPhT--T13/205
 }
\rule{1cm}{0cm}\vspace{19.5cm}\\
\fi

In these proceedings we present results from a recent
  calculation for the production of a $W$ boson in conjunction with
  five jets at next-to-leading order in perturbative QCD. We also use
  results at lower multiplicities to extrapolate the
  cross section to the same process with six jets. }

\FullConference{XXI International Workshop on Deep-Inelastic Scattering and Related Subject -DIS2013,\\
		22-26 April 2013\\
		Marseilles,France}

\begin{document}
\section{Introduction}
The production of a vector boson in conjunction with jets is an
important benchmark process for experiments at hadron
colliders. Reliable predictions for these processes are needed both as
a test of our understanding of Standard-Model measurements
and because they represent important backgrounds to many
searches for physics beyond the Standard Model (BSM). The study of the
newly discovered Higgs-like boson~\cite{CMSHiggs,AtlasHiggs}
 and searches for BSM signals requires
good modeling of high-multiplicity final states. The same holds true
for Standard-Model processes such as top-pair and di-boson production,
for which 
$W$+multijet processes represent
an important background. Because the neutrino from the leptonic
decay of the $W$ boson escapes the dectector, $W$+jets processes
represent an irreducible background for many searches involving
missing transverse energy.

Next-to-leading order (NLO) in the strong coupling constant is the lowest order at which reliable quantitative
predictions for cross sections are possible.  Theoretical predictions
for the
production of a $W$ boson in assoctiation with up to two jets 
at NLO have been available for a long 
time~\cite{NLOW1jet,MCFM,FernandoWjetpapers}.  At higher multiplicities,
predictions have become available only in recent years:
the so-called unitarity method (for a review, see
Refs.~\cite{ItaReview,EKMZReview}) has been used successfully for
$W$-boson production in association with
three~\cite{W3j}, four~\cite{W4j} and most recently five jets~\cite{W5j}.

Vector boson production has been measured in association with up to
four jets at the LHC with a center of mass energy of $7$~TeV
\cite{AtlasWjets,CMSWjets} and good agreement was found with
NLO predictions.
    
\section{W+jets at NLO}
\begin{figure*}[h]
\includegraphics[clip,scale=0.5]{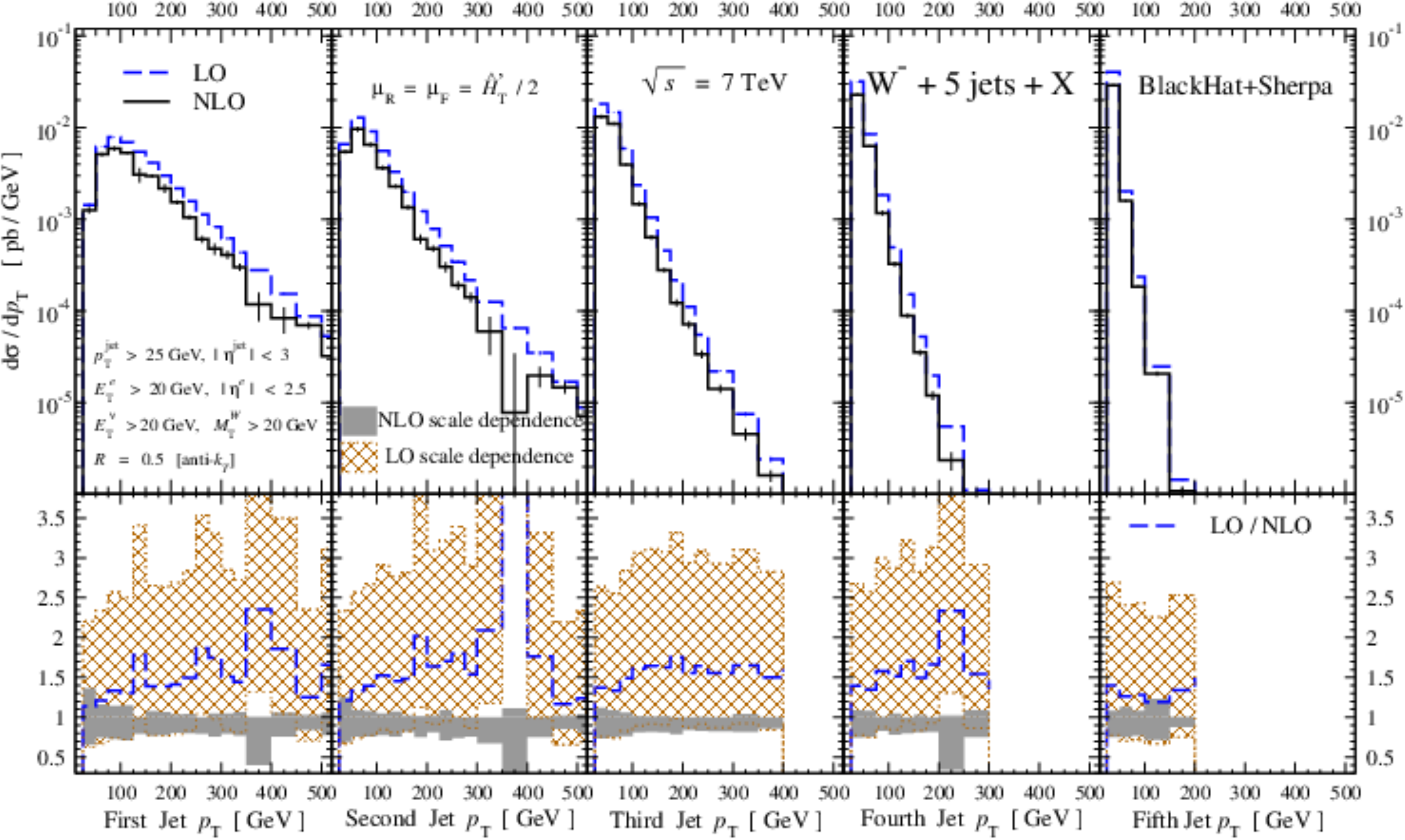}
\caption{The $\pT$ distributions of the leading five jets in
  \Wmjjjjj{}-jet production at the LHC at $\sqrt{s}=7$~TeV.  In the
  upper panels, the NLO predictions are shown as solid (black) lines,
  while the LO predictions are shown as dashed (blue) lines.
  The lower panels show the
  predictions for the LO distribution and scale-dependence bands
  normalized to the NLO prediction (at the scale $\mu=\HTpartonic'/2$).
The LO distribution is the dashed (blue) line, and the 
scale-dependence bands are shaded (gray) for NLO and cross-hatched
  (brown) for LO.  
}
\label{Wm5ptFigure}
\end{figure*}

We use \SHERPA~\cite{Sherpa} in
association with the virtual matrix elements from
\BlackHat~\cite{BlackHat}
to obtain our NLO predictions. We apply the jet and lepton cuts given in
Ref.~\cite{W5j}. The results for the total cross sections 
for \Wj-jet through \Wjjjjj-jet production are given in
table~1 of \cite{W5j}.
%
In Figure~\ref{Wm5ptFigure} we present the transverse momentum
distribution of the leading five jets. The striking feature is the
reduction of the scale variation obtained by going from LO to NLO. One
can also see that the radiative corrections affect not only the
overall normalization but also the shape of distributions.

Given that cross sections $\sigma_n$ are now available for a large
number of jet multiplicities $n$, one can investigate several ratios
involving these processes, which are interesting because some sources
of uncertainties largely cancel, both on the experimental and
theoretical side. The jet production and charge asymmetry ratios are defined as
\[
R^{W^\pm}_{n/(n-1)}=\frac{\sigma(W^\pm+n\;\mbox{jets})}{\sigma(W^\pm+(n-1)\;\mbox{jets})}\;,\qquad R_{W^+/W^-}=\frac{\sigma(W^++n\;\mbox{jets})}{\sigma(W^-+n\; \mbox{jets})}\;,
\]
repectively. Excluding the case $n=2$ where new partonic channels open in the
denominator, both ratios can be consistently described using a linear
interpolation, both at LO and NLO.

We collect the coefficients of the interpolation
$R=an+b$ for both ratios in the following table.
\begin{center}
\centering
\begin{tabular}{||c||c|c||c|c||}
\hline
 &  $a_{\rm{LO}}$  & $b_{\rm{LO}}$  &$a_{\rm{NLO}}$  & $b_{\rm{NLO}}$   \\  \hline
$R^{W^+}_{n/(n-1)}$ & $-0.0177 \pm 0.0004 $ & $0.320 \pm 0.002 $ & $ -0.009 \pm 0.003$ & $ 0.263 \pm 0.009$ \\ \hline
$R^{W^-}_{n/(n-1)}$ & $-0.0165 \pm 0.0005$  & $0.301 \pm 0.002$ & $0.009 \pm 0.002$  &  $ 0.248 \pm 0.008$ \\ \hline
$R_{W^+/W^-} $ & $0.102 \pm 0.002$  &$1.347 \pm 0.006$ & $0.11 \pm 0.01$  & $ 1.27 \pm 0.03$ \\ \hline
\end{tabular}
\end{center}

\begin{figure}[h]
\includegraphics[scale=0.38]{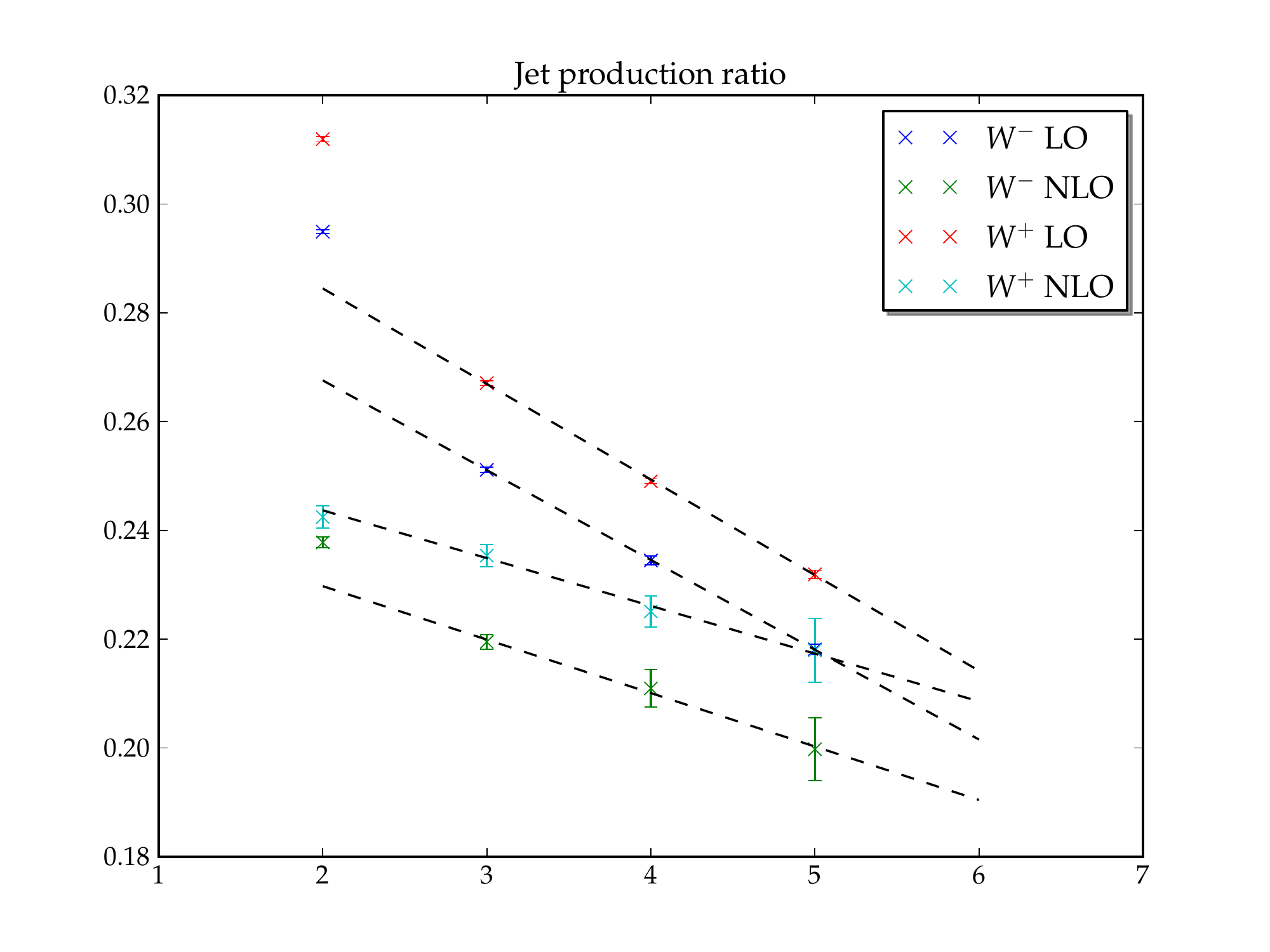}
\includegraphics[scale=0.38]{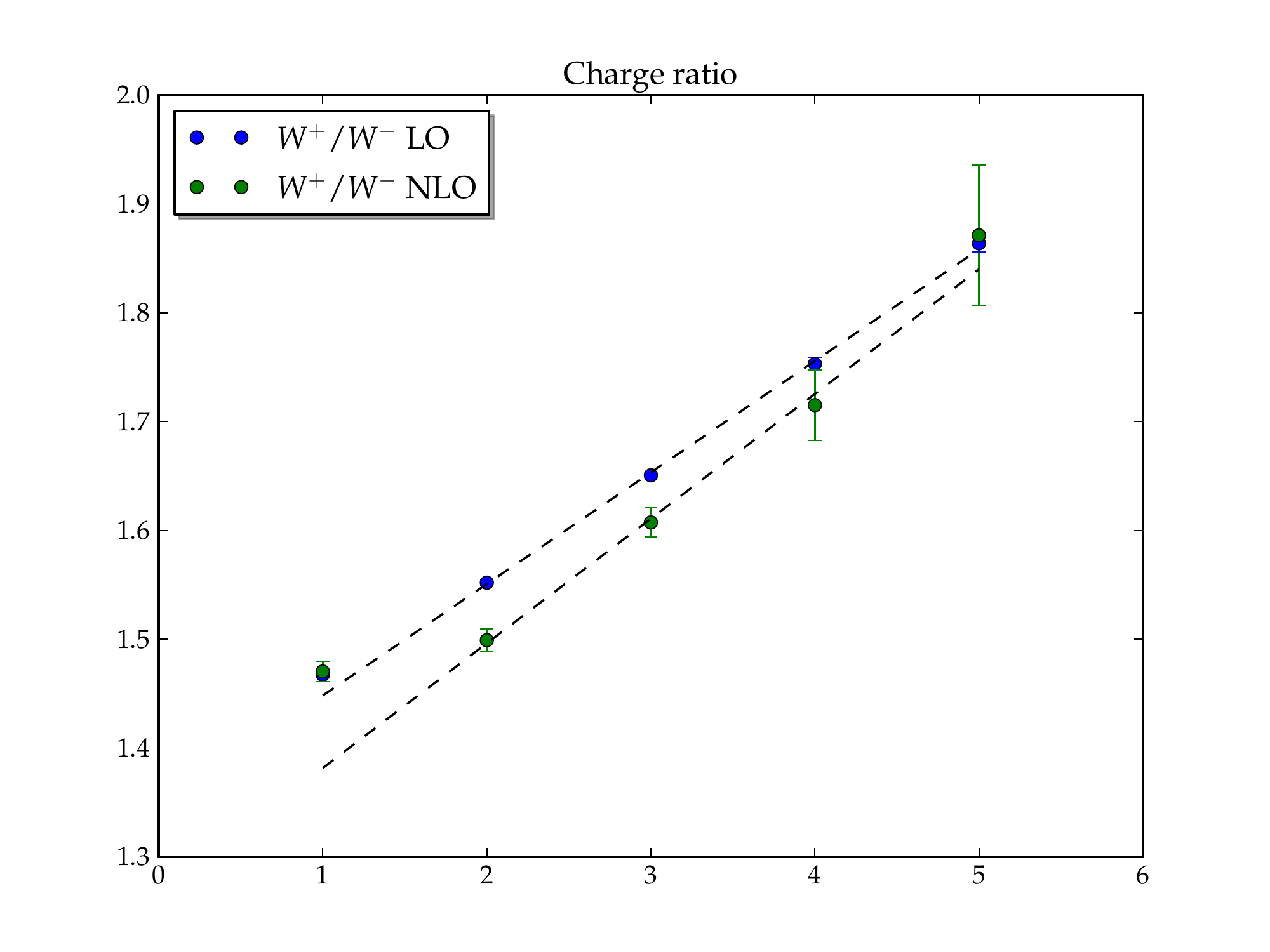}
\caption{Linear fits for the jet production and charge asymmetry
  ratios.}
\label{fig:interpolation}
\end{figure}

\begin{figure}[h]
\includegraphics[scale=0.45]{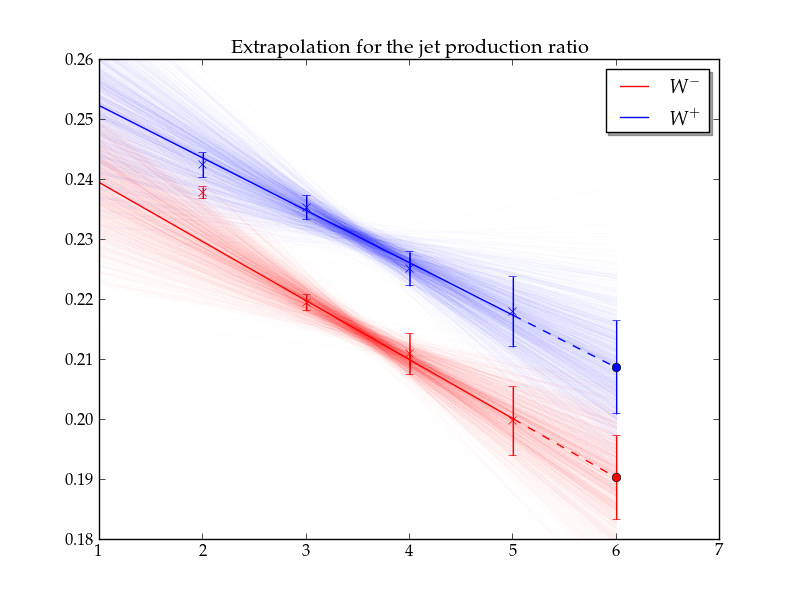}
\caption{Extrapolation of the jet production ratio. The error of the
  extrapolation is estimated from the variance of a synthetic set of
  linear extrapolations based on synthetic input data distributed in a
  gaussian around the central values with the width given by the
  statistical error. Each line represents one of these
  fits.}\label{fig:extrapolation}
\end{figure}

Using this linear regression one can extrapolate to obtain a prediction
for the \Wjjjjjj-jet cross section:
\begin{eqnarray*}
&&\sigma^{\mbox{extr}}(W^++6\;\mbox{jets})=0.30\pm 0.03\;\mbox{pb}\,,\\
&&\sigma^{\mbox{extr}}(W^-+6\;\mbox{jets})=0.15\pm 0.01\;\mbox{pb}\,.
\end{eqnarray*}
Englert {\it et al.\/} recently investigated~\cite{Gerwick} such an
extrapolation using jet calculus methods, and found it to be a good approximation when the jets
are required to have the same minimum transverse momenta. In our extrapolation, the errors
are computed from the variance of a large set of synthetic data
distributed in Gaussians around the central values of the cross
sections, with widths set to the corresponding
Monte-Carlo statistical errors.  The procedure is
illustrated in Fig~\ref{fig:extrapolation}.

\section{Conclusions}
We have presented results from a NLO calculation for the production
of a $W$ boson in association with up to five jets. We used these
results to provide an extrapolation for the cross section of the 
corresponding
process with six jets.

\section*{Acknowledgments}
This research was supported by the US Department of Energy under
contracts DE--FG03--91ER40662 and DE--AC02--76SF00515.  DAK's research
is supported by the European Research Council under Advanced
Investigator Grant ERC--AdG--228301.  DM's work was supported by the
Research Executive Agency (REA) of the European Union under the Grant
Agreement number PITN--GA--2010--264564 (LHCPhenoNet). The work of KJO
and SH was partly supported by a grant from the US LHC Theory
Initiative through NSF contract PHY--0705682.  This research used
resources of Academic Technology Services at UCLA, and of the National
Energy Research Scientific Computing Center, which is supported by the
Office of Science of the U.S. Department of Energy under Contract
No.~DE--AC02--05CH11231.

\end{document}